\documentclass[sigconf, nonacm]{acmart}

\newcommand\vldbdoi{XX.XX/XXX.XX}
\newcommand\vldbpages{XXX-XXX}
\newcommand\vldbvolume{14}
\newcommand\vldbissue{1}
\newcommand\vldbyear{2020}
\newcommand\vldbauthors{\authors}
\newcommand\vldbtitle{\shorttitle} 
\newcommand\vldbavailabilityurl{https://github.com/bytedance/videx}
\newcommand\vldbpagestyle{plain} 

\usepackage{booktabs}   
\usepackage{subcaption} 
\usepackage{comment} 
\usepackage{algorithm}
\usepackage{algpseudocode}
\usepackage{url}      
\usepackage{hyperref} 
\usepackage{booktabs} 
\usepackage{tabularx} 
\usepackage{xcolor} 
\usepackage{caption}
\usepackage{marvosym}
\usepackage{balance}

\usepackage{listings}
\usepackage{color}
\usepackage{pifont}

\definecolor{codegreen}{rgb}{0,0.6,0}
\definecolor{codegray}{rgb}{0.5,0.5,0.5}
\definecolor{codepurple}{rgb}{0.58,0,0.82}
\definecolor{backcolour}{rgb}{0.95,0.95,0.92}

\newcommand{\cmark}{\textcolor{green}{\ding{51}}}  
\newcommand{\xmark}{\textcolor{red}{\ding{55}}}    
\newcommand{\partialmark}{\textcolor{orange}{\ding{57}}}  

\lstdefinestyle{mystyle}{
    backgroundcolor=\color{backcolour},   
    commentstyle=\color{codegreen},
    keywordstyle=\color{magenta},
    numberstyle=\tiny\color{codegray},
    stringstyle=\color{codepurple},
    basicstyle=\footnotesize,
    breakatwhitespace=false,         
    breaklines=true,                 
    captionpos=b,                    
    keepspaces=true,                 
    numbers=left,                    
    numbersep=5pt,                  
    showspaces=false,                
    showstringspaces=false,
    showtabs=false,                  
    tabsize=2
}

\author{Rong Kang, Shuai Wang, Tieying Zhang$^*$, Xianghong Xu, Linhui Xu, Zhimin Liang, \mbox{Lei Zhang, Rui Shi, Jianjun Chen}}
\affiliation{%
  \institution{ByteDance}
  \country{}
}
\affiliation{
  {\{kangrong.cn,wangshuai.will,tieying.zhang,xuxianghong,xulinhui,liangzhimin,zhanglei.michael,\\shirui,jianjun.chen\}@bytedance.com}
  \country{}
}

\begin{document}
\title{
  VIDEX: A Disaggregated and Extensible Virtual Index for the Cloud and AI Era
  }


\begin{abstract}
Virtual index, also known as hypothetical indexes, play a crucial role in database query optimization. However, with the rapid advancement of cloud computing and AI-driven models for database optimization, traditional virtual index approaches face significant challenges. Cloud-native environments often prohibit direct conducting query optimization process on production databases due to stability requirements and data privacy concerns. Moreover, while AI models show promising progress, their integration with database systems poses challenges in system complexity, inference acceleration, and model hot updates. In this paper, we present VIDEX, a three-layer disaggregated architecture that decouples database instances, the virtual index optimizer, and algorithm services, providing standardized interfaces for AI model integration. Users can configure VIDEX by either collecting production statistics or by loading from a prepared file; this setup allows for high-accurate what-if analyses based on virtual indexes, achieving query plans that are identical to those of the production instance. Additionally, users can freely integrate new AI-driven algorithms into VIDEX. VIDEX has been successfully deployed at ByteDance, serving thousands of MySQL instances daily and over millions of SQL queries for index optimization tasks.

\end{abstract}

\maketitle

\pagestyle{\vldbpagestyle}
\begingroup\small\noindent\raggedright\textbf{PVLDB Reference Format:}\\
\vldbauthors. \vldbtitle. PVLDB, \vldbvolume(\vldbissue): \vldbpages, \vldbyear.\\
\href{https://doi.org/\vldbdoi}{doi:\vldbdoi}
\endgroup
\begingroup
\renewcommand\thefootnote{}\footnote{\noindent
$^*$Tieying Zhang is the corresponding author.\\
- This paper has been accepted by VLDB 2025.\\
This work is licensed under the Creative Commons BY-NC-ND 4.0 International License. Visit \url{https://creativecommons.org/licenses/by-nc-nd/4.0/} to view a copy of this license. For any use beyond those covered by this license, obtain permission by emailing \href{mailto:info@vldb.org}{info@vldb.org}. Copyright is held by the owner/author(s). Publication rights licensed to the VLDB Endowment. \\
\raggedright Proceedings of the VLDB Endowment, Vol. \vldbvolume, No. \vldbissue\ %
ISSN 2150-8097. \\
\href{https://doi.org/\vldbdoi}{doi:\vldbdoi} \\
}\addtocounter{footnote}{-1}\endgroup

\ifdefempty{\vldbavailabilityurl}{}{
\vspace{.3cm}
\begingroup\small\noindent\raggedright\textbf{PVLDB Artifact Availability:}\\
The source code, data, and/or other artifacts have been made available at \url{\vldbavailabilityurl}.
\endgroup
}

\section{Introduction}

Virtual index mechanisms facilitate hypothetical ``what-if'' analysis in a cost-efficient manner for query optimization evaluations, allowing database administrators to simulate the performance impact of potential indexes without actually creating them. This capability is implemented in most modern Database Management Systems (DBMSs). With the rapid advancement of cloud databases and AI-driven optimization techniques, two key challenges have emerged for database optimization systems: (1) Cloud-native environments require strict isolation between production instances and optimization components to ensure stability and data security; (2) AI-driven optimization approaches need flexible and standardized interfaces to integrate effectively with database systems. However, current virtual index implementations in popular DBMSs cannot address these requirements simultaneously, limiting their effectiveness in cloud-native environments and AI-driven optimization scenarios.

Commercial DBMSs, such as Oracle and SQL Server, are closed-source, which complicates the implementation of extensions for AI model interfaces in database optimization. PostgreSQL offers an open-source virtual index implementation, HypoPG\footnote{\url{https://github.com/HypoPG/hypopg}}, yet this solution is tightly integrated with the database instance, posing significant challenges to disaggregating optimization operations from production environments and incorporating AI interfaces. Furthermore, MySQL, one of the most prevalent open-source DBMSs, currently lacks a publicly available virtual index implementation. Consequently, there exists a pressing need within the database community to address these limitations to facilitate advanced optimization techniques in cloud-native environments.

\begin{table*}[tp]
  \centering
  \caption{Comparison of VIDEX for MySQL with {popular DBMSs} regarding key features, where \; \cmark \; represents satisfaction, \xmark \; indicates dissatisfaction, and \; \partialmark \; denotes partial satisfaction.}
  \vspace{-7pt} 
  \label{tab:feature_comparison}
  \begin{tabular}{lccccc}
    \toprule
  \textbf{Feature} & \textbf{Oracle} & \textbf{SQL Server} & \textbf{PostgreSQL} & \textbf{MySQL} & \textbf{VIDEX for MySQL} \\
  \midrule
  Disaggregated from Production & \cmark & \cmark & \xmark & \xmark & \cmark \\
  Enable Virtual Index & \cmark  & \cmark & \cmark  & \xmark & \cmark \\
  Extensible Model Interfaces & \xmark & \xmark & \partialmark & \xmark & \cmark \\
  Open-Source Availability & \xmark & \xmark & \cmark & \cmark & \cmark \\
  \bottomrule
  \end{tabular}
  \vspace{-7pt} 
\end{table*}

To this end, we introduce VIDEX\footnote{A demonstration video of VIDEX is available at: \url{https://youtu.be/Cm5O61kXQ_c}}, a \textbf{VI}rtual index engine with \textbf{D}isaggregated and \textbf{EX}tensible properties, tailored for addressing database optimization challenges in the cloud and AI era.
VIDEX presents a universal three-layer architecture applicable to various database systems, while our implementation specifically targets MySQL to demonstrate its effectiveness and simultaneously address the current absence of virtual index capabilities in this widely-used DBMS.  
We present these layers as follows: (1) The production database layer manages the online databases and maintains the statistical metadata. We demonstrate how the maintained statistical data enables secure query optimization operations by eliminating the need to access source data, thus significantly enhancing both security and scalability in cloud environments. (2) The VIDEX-optimizer layer functions as a ``what-if'' analysis tool using the statistical metadata. We demonstrate that it can generate query plans identical to those of the original online instances, thereby significantly reducing optimization overhead. (3) The VIDEX-Statistic-Server layer facilitates the integration and application of AI-driven database optimization models. We demonstrate that it can flexibly invoke different user-defined models for query optimization, such as estimation models for cardinality or number of distinct values (NDV), thus substantially accelerating the adoption of AI-driven approaches in database optimization.

We conducted a comparative analysis of popular DBMSs and VIDEX for MySQL features, evaluating their practicality in the era of cloud and AI. First, disaggregation from production databases indicates whether the system can conduct query optimization using only statistical replicas without accessing the source data. Oracle and SQL Server support this feature, whereas MySQL does not. Although PostgreSQL can retrieve most statistics from the online database, certain statistics can only be computed from the source data\footnote{\url{https://www.postgresql.org/docs/current/catalog-pg-statistic.html}}. Second, regarding virtual index support, MySQL is the only system among those evaluated that lacks this functionality. Third, the extensible model interfaces criterion assesses whether systems provide flexible interfaces for integrating AI-driven query optimization models. Among the evaluated DBMSs, only PostgreSQL offers limited interfaces for AI models, and these are not available 
on-demand~\cite{han_VLDB_CardEst_2021}. 
Finally, open-source availability indicates whether users can freely modify and extend the system's features. Table~\ref{tab:feature_comparison} presents a comparison of key features between VIDEX for MySQL and several popular DBMSs, highlighting the distinct advantages of VIDEX for MySQL.

VIDEX for MySQL has been deployed in ByteDance's large-scale production environments, optimizing thousands of MySQL instances and hundreds of thousands of SQL templates daily, where its reliability and practicality have been validated in an enterprise setting. VIDEX for MySQL has been open-sourced\footnote{https://github.com/bytedance/videx}; we hope it can deliver more efficient database optimization techniques and boost AI-driven optimization applications deployed into database production environments.


\section{System Design}

\begin{figure}[h]
  \centering
  \includegraphics[width=0.98\linewidth]{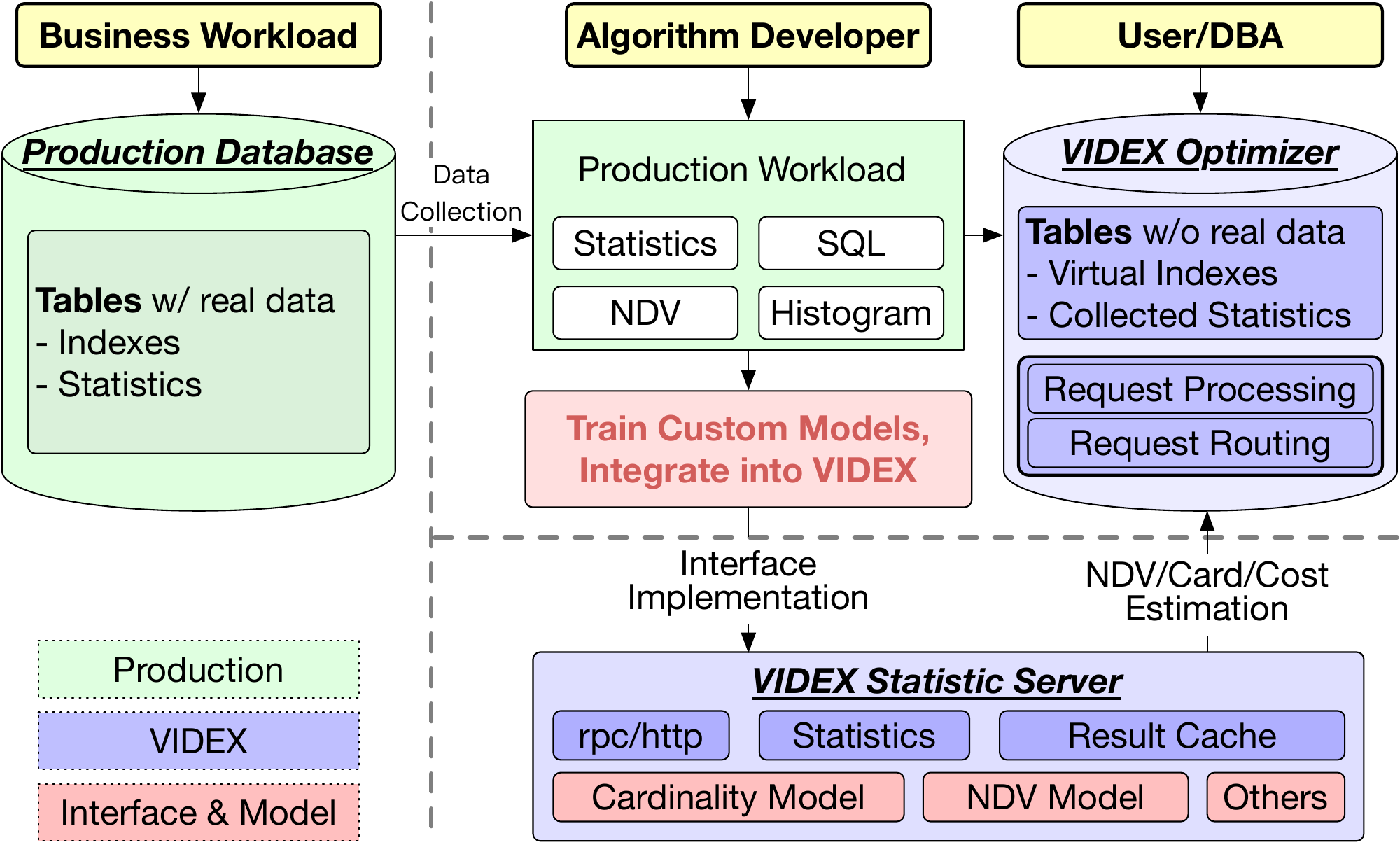}
  \caption{Three-layer, disaggregated VIDEX architecture}
  \label{fig:mysql-implementation}
  \vspace{-15pt} 
\end{figure}

\subsection{Architecture}

Figure ~\ref{fig:mysql-implementation} illustrates the three-layer disaggregated architecture design of VIDEX, which consists of three major components:

\noindent\textbf{Production database}: The production database manages large volumes of real data and typically maintains several existing indexes. The instance maintains both schema and statistical data. Schema information defines the table structure, indexes, and integrity constraints. Statistical metadata includes table rows, disk size, page numbers, histograms, and NDV. This statistical data may be automatically updated in real-time during database operations or gathered when users initiate collection processes. Each type of database employs a specified cost model within its query optimizer. Upon receiving a query, the database synthesizes a query plan using schema information, statistical metadata, and system configuration (such as \texttt{io\_block\_read\_cost} in MySQL).

\noindent\textbf{VIDEX-Optimizer}: The VIDEX-Optimizer is designed to simulate the query optimization process of production databases without requiring access to the actual data. However, achieving accurate simulation poses two major challenges: 

  \textbf{(1) Optimizer Logic Replication}: Database optimization tasks commonly use external third-party query optimizers, such as Apache Calcite~\cite{calcite}, to simulate database optimizers like MySQL and PostgreSQL~\cite{llm_r2,ali_das_sql}. However, third-party optimizers often do not perfectly reproduce the target production instances due to differing optimization rules and cost models, which can adversely affect the accuracy of what-if analyses.

  \textbf{(2) Statistical Data Replication}: The statistical information exerts a decisive influence on making query plans. We aim to create validation instances in an extremely lightweight manner, by synchronizing statistical information without creating data replicas. Although some commercial databases enable flawless replication of statistics from online production instances to offline validation instances, it remains a non-trivial task for many popular databases, such as MySQL. Currently, optimization service providers often create replicas of online databases for experimentation, but the synchronization of substantial data incurs additional costs.

We design the VIDEX-Optimizer component to address these challenges effectively. The VIDEX-Optimizer reutilizes the native optimizer of the target database and intercepts its requests for underlying statistical information. During its operation, it constructs virtual tables, referred to as ``VIDEX Tables,'' with schemas identical to those of the production instance. When SQL queries are directed towards the VIDEX-Optimizer, it sequentially performs logical and physical optimizations, both of which may require statistical information. For simpler statistics, such as table row counts, the VIDEX-Optimizer directly injects pre-collected statistics. For more complex requests, like cardinality estimation, the optimizer forwards these requests to the disaggregated VIDEX-Statistic-Server, which will be introduced subsequently. 

By decoupling query optimization from the production instance, VIDEX ensures a fully isolated environment for offline validation. This separation not only enhances stability in cloud-native scenarios but also permits database owners to analyze data using sanitized and audited statistics, thereby preserving data privacy and preventing the leakage of sensitive information.

\noindent\textbf{VIDEX-Statistic-Server}: VIDEX decouples complex, precision-sensitive estimation requests from the VIDEX-Optimizer, such as cardinality estimation, which are particularly suitable for AI optimization. These requests are forwarded to a dedicated algorithmic service, the VIDEX-Statistic-Server, via REST or RPC protocols. The VIDEX-Statistic-Server stores data collected from the production environment, including statistical data and historical queries, which may be utilized by query-driven methods~\cite{han_VLDB_CardEst_2021}. Based on the above data, the integrated AI algorithms can provide more accurate estimations. The VIDEX-Statistic-Server also supports response and model caching because we observe that different SQL queries often request repeated cardinality results, while some generalized pre-trained models only need to be loaded once to respond to multiple requests from different tables.

The design of disaggregated statistic server brings two benefits:

    \textbf{(1) Customizable AI-Driven GPU Optimization:} Decoupling into a separate service enables tailored optimizations of AI models independent of database updates. By deploying the VIDEX-Statistic-Server on GPU clusters, GPU acceleration is fully leveraged, allowing AI algorithms to evolve separately from database version upgrades and facilitating re-training and hot-updating of AI models.

    \textbf{(2) Extensible Algorithm Framework and Flexibility:} The VIDEX-Statistic-Server simplifies integration by offering standardized interfaces that enable easy algorithmic insertions, enhancing user-driven customization. Users can utilize data such as statistics, histograms, and historical queries collected by VIDEX for efficient model training and rapid deployment. Further details on these interfaces will be introduced subsequently.

\subsection{Workflow and Deployment}

When a user initiates an analytical task for a specific production database, VIDEX first collects the necessary metadata, including schema, system variables, and statistical data, either through authorized production APIs or from a user-prepared metadata file. Subsequently, a VIDEX optimizer instance is chosen to create schema-identical, dataless tables. The collected metadata is then imported into an assigned VIDEX Statistics Server instance. After these preparations, users can connect to their optimizer instance to conduct index modification operations and use SQL EXPLAIN to generate query plans highly similar to those in the production environment.

The VIDEX Optimizer is deployed as a standalone service, providing what-if analysis for large-scale production instances. Given that the VIDEX Optimizer requires only schema and statistical information, it is extremely lightweight. Consequently, a single VIDEX Optimizer instance can manage hundreds of analytical tasks from various business workloads. Furthermore, by leveraging the elasticity of cloud-native environments, the VIDEX Optimizer can be dynamically scaled to accommodate an increasing number of analytical tasks.

The VIDEX Statistics Server is also deployed as a standalone service and handles requests from the VIDEX Optimizer. All requests generated by the same task are routed to the same statistics server instance. If multiple tasks correspond to the same database, requests are routed to the instance that has already loaded the relevant statistical data and models, thus minimizing redundant data imports and model loads.

\subsection{Algorithm Interface}

We provide easy-to-understand and standard interfaces for custom algorithm integration. As shown in Listing~\ref{code:videx-interface}, VIDEX defines two core interfaces: \texttt{cardinality} estimation and \texttt{NDV} estimation. The cardinality interface takes structured range conditions as input and returns the estimated row count, while the NDV interface accepts the column name list, returning the estimated NDV. This clean abstraction enables researchers to focus on their estimation algorithms without the need to navigate complex database internals. Utilizing this framework, developers can seamlessly integrate various heuristic or machine learning-based approaches.

\begin{figure}[t]
  \begin{lstlisting}[language=Python, 
    caption={VIDEX Statistic Server Interfaces}, label=code:videx-interface]
class RangeCond:
  """Represents a single-column range condition"""
  col_name                    # Column name
  data_type                   # Data type
  min_value, max_value        # None, or boundary value
  min_operator, max_operator  # "<", "<=", ">", ">="

class VidexModelBase(ABC):
  def __init__(self, full_table_stats, model_path=None):
    self.full_table_stats = full_table_stats
    self.model_path = model_path

  @abstractmethod
  def cardinality(self, range_cond: List[RangeCond]):
    """Estimates number of rows matching conditions"""
    pass

  @abstractmethod
  def ndv(self, column_list: List[str]):
    """Estimates number of distinct values"""
    pass
\end{lstlisting}
\vspace{-20pt} 
\end{figure}

\begin{figure*}[htp]
  \centering
  \includegraphics[width=0.8\linewidth]{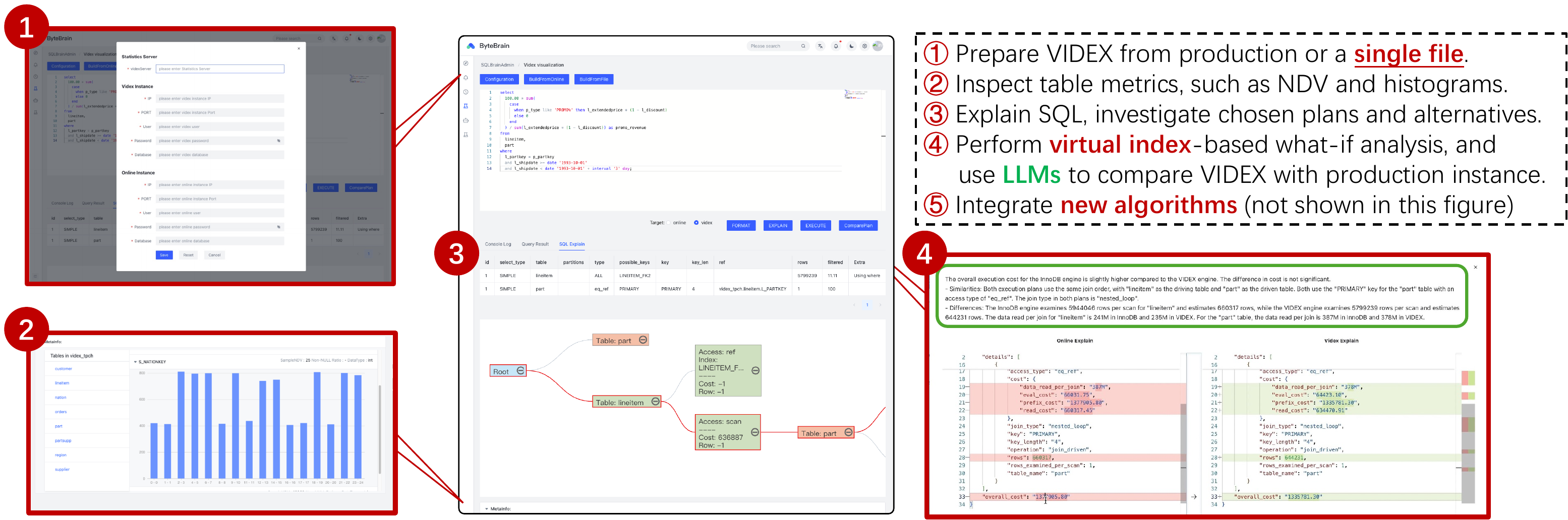}
  \caption{
    The illustration of the VIDEX Workflow.
    }
    \vspace{-10pt}
  \label{fig:videx-demo-scenario}
\end{figure*}

\section{Implementation}
\label{sec:implementation}

\textbf{VIDEX-Optimizer:} Since MySQL lacks a native virtual index mechanism and restricts schema modifications, we implemented the VIDEX-Optimizer with an underlying layer named VIDEX-Engine. This engine interacts with MySQL's query optimizer and index operations. We meticulously reviewed over 90 engine interfaces and implemented essential interfaces for query optimization and index management. The VIDEX-Engine formats requests related to cardinality and NDV estimation in a user-friendly manner and forwards them to the VIDEX-Statistic-Server

\textbf{VIDEX-Statistic-Server:} We developed two solutions for metadata collection and cost estimation.
\textbf{1). Statistic Fetching}: This method leverages MySQL's system tables to gather basic statistics data. 
It collects single-column distinct values (NDV) and histograms via ``UPDATE HISTOGRAM'' and ``SELECT COUNT DISTINCT'', and leverages the column independence assumption to estimate multi-column join statistics.
\textbf{2). Data Sampling}: This method employs authorized sampling APIs to collect a subset of data (up to $10^5$ rows). It generates mutli-column histograms to estimate the joint cardinality from sampling data, and employs an pre-trained AI model AdaNDV~\cite{xu2025adandv} to estimate the joint NDV.

Notably, VIDEX can be engineered to extend to other databases such as PostgreSQL. VIDEX-Optimizer can utilize PostgreSQL's flexible hook mechanism and rich plugins, while the VIDEX-Statistic-Server is database-agnostic and interfaces with various database optimizers. This will be our future work.

\section{Experiments}
We assume that the NDV and cardinality algorithms can produce results identical to MySQL with full data, and evaluate VIDEX's ability to generate accurate query plans under this premise. As shown in Table~\ref{tab:experiments}, across all three benchmarks (TPC-H, TPC-H-Skew, and JOB), the VIDEX query plans perfectly match the expected results, including identical join orders and index selections. The row count estimates closely match native MySQL, with average q-error per plan operator (max ratio between VIDEX and native MySQL estimated rows) less than 1.1x. Furthermore, even without perfect statistical information, our solutions in Section~\ref{sec:implementation} maintain high simulation fidelity for workloads with independent correlations and uniform distributions, as demonstrated in TPC-H and numerous ByteDance production instances.

\begin{table}[htp]
  \centering
  \caption{Simulation Accuracy: VIDEX vs. Native MySQL}
  \vspace{-4pt}
  \label{tab:experiments}
  \small
  \begin{tabular}{lcccr}
    \toprule
    \textbf{Benchmark} & \textbf{\#SQL} & \textbf{Join Order} & \textbf{Index Select} & \textbf{Row Q-Error} \\
    \midrule
    TPC-H      & 22  & Identical & Identical & 1.08 \\
    TPC-H-Skew & 22  & Identical & Identical & 1.05 \\
    JOB        & 103 & Identical & Identical & 1.09 \\
    \bottomrule
  \end{tabular}
  \vspace{-12pt}
\end{table}

\section{Demonstration Scenario}

This section presents a scenario to demonstrate VIDEX's capabilities in disaggregated architecture, accurate MySQL query plan simulation, virtual index-based what-if analysis, and framework extensibility, as shown in Figure~\ref{fig:videx-demo-scenario}. Additionally, our Web GUI enables detailed statistical analysis and supports LLM-enhanced differential analysis between VIDEX and the target production instance.

\textbf{\ding{172} $\sim$ \ding{173} Prepare VIDEX and Inspect Table Metrics}. Initially, we enter the connection details for three components: a production MySQL instance loaded with the TPC-H dataset, a VIDEX-Optimizer instance, and a VIDEX-Statistic server. VIDEX can gather metadata and statistical data either by directly connecting to the production instance or by uploading a file. Subsequently, VIDEX create tables within the VIDEX-Optimizer and populates the VIDEX-Statistic Server with metadata. We can then inspect table metrics, such as NDV and histograms.

\textbf{\ding{174} SQL explain and detailed analysis}. We perform an SQL EXPLAIN on TPC-H Q14, subsequently reviewing the EXPLAIN results. The Web GUI additionally offers an interactive query plan in tree format, enabling the comparison of the selected query plan (marked with a red edge) with alternative candidates.

\textbf{\ding{175} What-if analysis via virtual index and compare with production database}. We add a virtual index via VIDEX and observe a significant reduction in costs. Subsequently, we create a real index on the production system, noting a notable decrease in execution time, as expected. Furthermore, the Web GUI facilitates comparisons between the query plans of VIDEX and the production instance, accentuating differences and employing LLMs for explanations. VIDEX consistently generates query plans that closely resemble those of the production system, thus demonstrating the high-precision simulation capabilities of its virtual index mechanism for MySQL.

\textbf{\ding{176} Conveniently add a new cardinality algorithm}. Finally, we illustrate the extensibility of the VIDEX algorithm framework by developing a Python implementation that includes a simple NDV and cardinality estimation function.

\bibliographystyle{ACM-Reference-Format}
\bibliography{videx_demo}

\end{document}